
\input harvmac

\hyphenation{anom-aly anom-alies coun-ter-term coun-ter-terms}
\def\inv{^{\raise.15ex\hbox{${\scriptscriptstyle -}$}\kern-.05em 1}}

\def\Dsl{\,\raise.15ex\hbox{/}\mkern-13.5mu D} 
\def\dsl{\raise.15ex\hbox{/}\kern-.57em\partial}

\def\lspace{\ifx\answ\bigans{}\else\qquad\fi}
\def\lbspace{\ifx\answ\bigans{}\else\hskip-.2in\fi} 
\def\boxeqn#1{\vcenter{\vbox{\hrule\hbox{\vrule\kern3pt\vbox{\kern3pt
	\hbox{${\displaystyle #1}$}\kern3pt}\kern3pt\vrule}\hrule}}}
\def\mbox#1#2{\vcenter{\hrule \hbox{\vrule height#2in
		\kern#1in \vrule} \hrule}}  
%

\def\om#1#2{\omega^{#1}{}_{#2}}

\def\darr#1{\raise1.5ex\hbox{$\leftrightarrow$}\mkern-16.5mu #1}

\def\roughly#1{\raise.3ex\hbox{$#1$\kern-.75em\lower1ex\hbox{$\sim$}}}

\def\square{\kern1pt\vbox{\hrule height 1.2pt\hbox{\vrule width 1.2pt\hskip 3pt
   \vbox{\vskip 6pt}\hskip 3pt\vrule width 0.6pt}\hrule height 0.6pt}\kern1pt}
\def\newpage{\vfill\eject}

\def\fun#1#2{\lower3.6pt\vbox{\baselineskip0pt\lineskip.9pt
  \ialign{$\mathsurround=0pt#1\hfil##\hfil$\crcr#2\crcr\sim\crcr}}}

\def\om{\omega}

\def\p{\Phi}

\def\phi{\Phi}

\def\part{\partial}
\def\ep{\epsilon}

\def\lta{\mathrel{\spose{\lower 3pt\hbox{$\mathchar"218$}}
     \raise 2.0pt\hbox{$\mathchar"13C$}}}
\def\gta{\mathrel{\spose{\lower 3pt\hbox{$\mathchar"218$}}
     \raise 2.0pt\hbox{$\mathchar"13E$}}}
\def\spose#1{\hbox to 0pt{#1\hss}}

\vsize=23.0truecm
\hsize=16.25truecm
\parskip=0.25truecm
\def\newpage{\vfill\eject}

\def\fun#1#2{\lower3.6pt\vbox{\baselineskip0pt\lineskip.9pt
  \ialign{$\mathsurround=0pt#1\hfil##\hfil$\crcr#2\crcr\sim\crcr}}}
\baselineskip=24truept

\def\mad{MAD\ }
%
%
$\,$

\centerline{\bf  Curvature and Flatness}
\centerline{\bf in a Brans-Dicke Universe}
\vskip 0.2truein
\vskip 0.1truein
\centerline{ {\bf Janna  J. Levin$^1$ and Katherine Freese$^2$}}
\centerline{\it $^1$Department of Physics, MIT}
\centerline{\it Cambridge, MA 02139}
\vskip 0.2truein
\centerline{\it $^2$Physics Department, University of Michigan}
\centerline{\it Ann Arbor, MI 48109}
\centerline{{\rm and} {\it Institute for Theoretical Physics}}
\centerline{\it Santa Barbara, CA 93106}

\centerline{\bf Abstract}

The evolution of a universe with Brans-Dicke  gravity
and nonzero curvature is investigated here.
We present the equations of motion and their solutions
during the radiation dominated era.
In a Friedman-Robertson-Walker cosmology we show explicitly that the three
possible values of curvature $\kappa=+1,0,-1$ divide the evolution
of the Brans-Dicke universe into dynamically distinct classes
just as for the standard model.  Subsequently
we discuss the flatness problem which exists in
Brans-Dicke gravity as it does in the standard model.
In addition, we demonstrate a
flatness problem in MAD Brans-Dicke gravity.
In general,
in any model that addresses
the horizon problem, including inflation, there are two components
to the flatness issue: i) at the Planck
epoch curvature gains importance, and ii) during accelerated expansion
curvature becomes less important and the universe
flattens.
In many cases the universe must be very flat
at the Planck scale in order for the accelerated epoch to be reached,
thus there can be a residual flatness problem.


\centerline{to be published in  {\it Nuclear Physics B},
submitted August 30, 1993}

\newpage

\vskip 20truept
\centerline{\bf Introduction}

In the Brans-Dicke theory of gravity,
 the constant  Planck mass of the Einstein theory is replaced with
a  massless scalar field.$^1$
As a result, the gravitational constant is not a fundamental
constant of the theory but instead, the strength of gravity
evolves dynamically.
Interest in alterations to Einstein gravity
has arisen in a variety of contexts.
The cosmological importance
of such theories has been investigated
in inflationary models
such as (hyper)-extended inflation$^{2}$ and
Starobinsky's cosmology$^{3}$.
Other cosmological implications
of modifying gravity have been indicated
in attempted
alternatives to inflation such as the MAD
prescription$^{4,5}$. In addition, some innovative theories
which try to reconcile particle
physics with gravity  lead to  low energy
theories which behave like the Brans-Dicke model.
For instance, higher dimensional theories or Kaluza-Klein
theories$^6$ can lead to a
dynamical Planck mass.

For a large part of this paper,
we describe
the evolution of a universe with Brans-Dicke
gravity and nonzero curvature.
We present the equations of motion and
their solutions$^{7}$ during the radiation dominated era.  We find the
evolution of the  scale factor, the temperature, and the Hubble
constant as a function of the changing
Planck mass rather than explicitly
as a function of time -- see Ref. [4]
for some discussion of explicit time
dependence for a flat Brans-Dicke cosmology.
Although these solutions were presented in Ref [4]
for a flat universe,
the case of nonzero curvature was given only
cursory  attention.
In this paper we study more thoroughly
the evolution of curved Brans-Dicke
cosmologies.

We begin by solving the equations
of motion for
general curvature.
As expected, for a Friedmann Robertson Walker (FRW)
metric, we'll see that Brans-Dicke models
can be split into three cases as in the standard model:
the three possible curvatures
$\kappa=+1,0,-1$ break the universe up into dynamically
distinct classes.
In the $\kappa=+1$ universe, the energy density in matter
exceeds the kinetic energy of the expansion. Eventually
the expansion will cease and the universe will collapse under the pull
of its own weight.  If $\kappa=-1$ the cosmology is open.
The energy density in matter is not sufficient to close the universe
and
it expands forever.
The critical case, separating these two is the flat cosmology,
$\kappa=0$
for which
there is just enough
 kinetic energy to escape
collapse.

Once we have built a picture of the large-scale
behavior of curved Brans-Dicke cosmologies we
can ask if these cosmologies
have a flatness problem.
We devote the latter half of the paper to a study
of flatness.
The Brans-Dicke cosmology by construction evolves adiabatically
and so, as we show, will have a flatness problem.
If the Planck mass were to couple directly to matter,
then the assumption of adiabaticity is unfounded.
It would be interesting in the future to investigate
this possibility.
Finally, at the end of the paper, we discuss flatness in the \mad
solution to the horizon and monopole problems; the MAD
proposal also
relies on a dynamical Planck mass such as occurs in scalar
theories of gravity.
(For a discussion of the limitations and future of this model
see Ref [8] and [9].)
We show that if the cosmic evolution is adiabatic,
as it is for the Brans-Dicke model,
then \mad  Brans-Dicke gravity cannot resolve the flatness problem.

Before proceeding, we introduce the flatness problem.
To begin we review this problem in the context of the standard model.
It appears that the universe has survived to  a temperature
of $T_o=2.74^o$ K and a ripe old age of 10-15 billion years.
The survival of our universe,
 in the context of
the standard hot big bang cosmology,
requires extraordinary values of some
otherwise arbitrary constants.
That is,
for our universe to survive with these conditions it
must be
that curvature does not completely dominate the cosmic evolution.
Yet, in the standard cosmology, the
universe
should quickly veer away from a flat appearance
unless
extraordinary initial
conditions are imposed
which render  the universe extremely close to flat at its inception.

Consider the standard model
Einstein equation in a Friedmann Robertson Walker (FRW)
cosmology
		\eqn\normal{H^2+\kappa/R^2=(8\pi/3M_o^2)\rho\ \ .}
The
curvature term in the
equation of motion \normal\ scales as $1/R^2$ while the
radiation density term scales as $\rho\sim
1/R^4$.
Consequently, as we look back in time, when the universe is very
small, the energy density dominates over curvature.
Initially, curvature is unimportant in determining the dynamics
of the scale factor and the universe looks roughly flat.
As $R$ grows, the curvature term should
quickly come to dominate in the determination
of the cosmological evolution.
The fact that the matter term is still
significant implies that the curvature
radius defined from
    \eqn\curvy{
      R_{\rm curv}={R(t)\over |\kappa|^{1/2}\  \  }   }
must
be greater than or comparable to the Hubble length $H^{-1}$;
      \eqn\curvbig{
        R_{\rm curv}\gta  H^{-1}\ \ .}
Multiplying both sides of  eqn \curvbig\
by the temperature $T$ and cubing we
have the condition that the entropy within a curvature
volume,
	\eqn\trythis{
	S_{\rm curv}=R^3_{\rm curv} T^3=
	R^3(t)T^3(t)|\kappa|^{-3/2}\equiv \bar S|\kappa|^{-3/2}
	\ \ ,
	}
must exceed
$H^{-3}_oT^{3}_o$, which is roughly
the entropy within a Hubble volume;
        \eqn\hm{
 	\bar S|\kappa|^{-3/2}
      \gta  H_o^{-3}T_o^3= \alpha_o^{-3/2}{M_o^3\over T_o^3}\ \ ,
      }
 with
$\alpha_o=\gamma(t_o)\eta_o=8\pi/3(\pi^2/30)g_*(t_o)\eta_o$
where
$\eta_o\sim 10^{4}-10^5$
is the
ratio today of the energy density in matter to
that in radiation.
Notice $\bar S (4/3)(\pi^2/30)g_S\equiv S$
where $S$ is the constant of motion and $g_S$ counts
the number of degrees of freedom contributing to the entropy.
The constant Planck mass of the Einstein theory is
$M_o=1.2\times 10^{19}$
GeV and the temperature of the cosmic background radiation
in units of GeV is
$T_o=2.3\times 10^{-13}$ GeV.
Then eqn \hm \
demands that
$\bar S|\kappa|^{-3/2}\gta  {10^{90}}$.
As long as the cosmic evolution is adiabatic, then
$\bar S$ and  $S_{\rm curv}$
are constant, up to factors of degrees of freedom.
Notice, if the universe is flat and $\kappa=0$ eqn
\hm \ is automatically satisfied.  If instead the universe
is created with $\kappa=\pm 1$, then  eqn
\hm \ tells
us that if the universe is to survive until today, with
the conditions we observe, then the otherwise arbitrary
constant entropy $\bar S$ must
have a monstrous value in excess of  roughly $10^{90}$.
Thus an extraordinary
value of an arbitrary constant of motion is required to
preserve  our cosmology.
The challenge is to explain the enormous value of this
otherwise arbitrary
constant.$^{10}$

In modifications to the standard model
which attempt to address the related horizon and
monopole problems, the flatness problem must
be reexamined.
As we'll see,  in these dynamical models
there are two components to the
flatness  issue:
1)  At some high temperature, the cosmology
undergoes an accelerated expansion, as happens
for instance when an
inflationary epoch begins.$^{11}$
During acceleration
curvature becomes less important and the universe
becomes flatter as we demonstrate below.
2)  Above the temperature at which acceleration ensues,
there is first an early epoch during which the universe decelerates
and curvature gains importance [unless, of course, the
accelerated expansion e.g. inflation takes
place at the Planck scale].

To see these two components to the flatness problem,
consider first dynamic solutions to the horizon problem.
One can express the causality condition
required to solve the horizon problem
in a simple way:
    \eqn\nut{{1 \over H_c R_c} \geq {1 \over H_o R_o}\ \ .}
(This equation is not the most general.
It holds only if the scale factor of the universe
behaves as a simple power law in time
before $t_c$ and during matter domination.
See also below eqn (51).)
If this equation is satisfied, our observable
universe today fits inside a causally connected
region at some early time $t_c$.
Note that this equation implies that
$\ddot R > 0$ for some
period between $t_c$ and the present.
The most successful model to date
that satisfies eqn \nut\ is inflation.
MAD models attempt to satisfy eqn \nut\
by replacing the potential domination in inflationary models with a
change in the behavior of gravity.
In any case, any model that satisfies this condition will
automatically make the universe flatter.
We can demonstrate this  by comparing the scales
$R_{\rm curv}^{-1}$  and
$H$:
  \eqn\bas{{ R_{\rm curv}^{-1}\over H}={|\kappa|^{1/2}\over \dot R }
\ \ .}
We have argued that
any dynamical model which solves the horizon
problem, will accelerate the cosmic expansion.
 As the universe accelerates,
$\dot R$ must in fact grow.
The importance
of curvature  will only
diminish as $\dot R$ grows, thus rendering the universe flatter.
Therefore, any dynamical model that satisfies eqn \nut\
inevitably makes the universe flatter.

However, there is a second component to the flatness
problem.   Starting at the
Planck time, before the onset of the
accelerating phase, the universe  decelerates
and curvature gains importance.
Again, we can see this from eqn \bas .
As $\dot R$ slows, the curvature term grows
in importance in determining the cosmic evolution.
One has to be cautious that the earliest era
during which curvature gains importance
does not
generate a serious
flatness problem.
For an adiabatic model, it is this early aspect of the flatness
problem which is not escaped.$^5$  An adiabatic \mad universe
therefore has a flatness problem as we will
show in detail toward the end of the paper.

Inflation generates a large value for $\bar S$ today
by dynamically producing entropy.
If inflation begins at a temperature $T_c = M_o$,
then the flatness problem is solved.
[To reiterate,
$M_o = 1.2\times 10^{19}$ GeV is the standard Planck mass.]
However, if inflation begins significantly
below the Planck scale, i.e. $T_c \ll M_o$,
and if the universe is closed,
then there is a residual, though less severe,
flatness problem.
In order for the temperature in a closed universe
to reach
$T_c$, the temperature at which inflation begins,
a large entropy  is required,
$\bar S\gta ( M_o/T_c)^3$,
as is shown in \S VIIB.
Unless inflation begins near the Planck scale, there will be a large
constraint on the entropy.
For instance if inflation begins near a temperature
of $T_c\sim 10^{14}$ GeV, then the entropy must
exceed $\bar S\gta 10^{15}$.
If the entropy is not at least this large, then the universe
collapses  before inflation begins.
In an open cosmology, the universe will tend away from flatness
by the time inflation begins (again, if $T_c < M_o$).
To correct for this, inflation
requires either (i) extra e-foldings of inflation
if initially $\bar S \sim 1$ or (ii) an initial value of
$\bar S\gta (M_o/T_c)^3$.
These claims about flatness are explained in detail in the paper.

\centerline{\bf II. Equations of motion and their Solutions}

In a scalar theory of gravity, such as that proposed by Brans and Dicke, the
Einstein action, $A_{\rm einst}=\int d^4x\sqrt{-g} \left({-M_o^2 /
16\pi}\right){\cal R}$  where $\cal R$ is the Ricci scalar
and $M_o= 10^{19}$GeV, is replaced by a
coupling between the Ricci scalar and some function of a scalar field $\psi$.
We will call the function of the scalar field $\p$ and note that
$<\p >\equiv m_{pl}^2$.  Thus the Planck mass, which dictates
the strength
of gravity, is determined dynamically by the expectation value of $\p$.
The $<>$ will be implicit in the rest of the paper.  The
action describing the theory is given by
    \eqn\je{A=\int d^4x\sqrt{-g}\left[-{\Phi(\psi)\over 16\pi }{\cal R}
        - {\omega \over \Phi}{\partial_{\mu}\Phi
        \partial^{\mu}\Phi
        \over 16\pi}+{\cal L}_{\rm m}\right ]\ \ , }
where we used the metric convention $(-,+,+,+)$, and ${\cal L}_{\rm m}$ is
the Lagrangian density for all the matter fields excluding the field
$\psi$.  The parameter  $\omega$ is
defined by $\omega=8\pi{\Phi \over  (\partial\Phi/\partial \psi)^2}$.
In this paper we consider
the original proposal of Brans and Dicke,
        \eqn\organ{\p ={2\pi\over \omega}\psi^2\ \ }
where $\om$
is a  constant parameter of the theory.
Notice that there is no direct coupling of the Planck
mass to ${\cal  L}_{\rm m}$.  As a consequence of this, the
universe evolves adiabatically so that $R\propto T^{-1}$
as we describe below.

Stationarizing this action
with respect to the metric gives the Einstein-like equation
    \eqn\likein{G_{\mu\nu}={8\pi\over \p}[T^{\rm m}_{\mu \nu}
     +T^{\p}_{\mu\nu}]\ \ ,}
where $T^{\rm m}_{\mu\nu}$ is the energy-momentum tensor
in all fields excluding the Brans-Dicke field and
$T^{\p}_{\mu\nu}$ is the energy-momentum tensor in the $\p$ field.
In a Friedman-Robertson-Walker cosmology \likein\
gives the equation of motion for the scale factor $R(t)$
  \eqn\two{H^2+{\kappa \over R^2}={8\pi \rho \over 3\Phi}
        -{\dot\Phi\over\Phi} H +{\omega \over 6}\left({\dot \Phi
        \over \Phi}\right)^2 }
where $\kappa=0, +1,$ or $-1$ while
$\rho$ is the energy density and
$p$ is the
pressure in all fields excluding the
$\psi$ field.
The principle of stationary action
with respect to the coordinate $\p$
gives
  \eqn\one{\ddot \Phi +3H\dot \Phi={8\pi\over  3+2\omega }(\rho-3p)
       \ \ . }

Conservation of energy-momentum in the $\p$ sector,
 $-8\pi T^{\mu \nu}_{\Phi \ ;\mu}=({\cal R}^{\mu \nu}
        -{1\over 2}g^{\mu\nu}{\cal R})\Phi_{;\mu}\ ,$
is equivalent to the equation of motion of \one.
Conservation of
energy-momentum in the matter
sector can be satisfied independently,
$T^{\mu \nu}_{\rm m \ ;\nu}=0 $.
In an isotropic and homogeneous universe the $\mu=0$
component of the matter conservation eqn  gives
$\dot \rho =-(\rho+p)3H$.
Consider the radiation
dominated era where $\rho=(\pi^2/30)g_*T^4$, $p=\rho/3$, and $g_*$
is the number of relativistic degrees of freedom in equilibrium.
Since conservation of energy-momentum in ordinary matter
does not involve
$\Phi$, we can deduce  that
the entropy per comoving
volume in ordinary matter, $S=(\rho +p)V/ T$, is conserved.
We  use the definition
    \eqn\sox{ \bar S=R^3T^3\ \ ,}
 where $S \simeq \bar S
(4/3)(\pi^2/30)g_S$. For practical purposes we can take $g_*=g_S$.

We present here the solutions to the equations of motion
during a radiation dominated era for
a Brans-Dicke theory with
general $\kappa$.  The flat ($\kappa=0$)
cosmology was described in detail in Ref [4]
while the $\kappa \ne 0$ cases were briefly described in
the appendix of that reference.  Here
the curved cosmologies are considered
in detail.
A flat cosmology is included as a particular case of
these solutions.
We
parameterize $R$, $T$, and thus $H$ by the
Brans-Dicke field $\p$.

The first integral of the $\p$ motion gives
  \eqn\again{\dot \p R^3=-C \ \ , \ \
{\rm also}\ \ \
H=-{\ddot \p\over 3\dot \p}\ \ \  .}
$C$ is an arbitrary constant of integration which can be
positive, negative, or zero.
Consider the case of $C$ identically zero.
Then the Planck mass is constant
and the cosmology imitates
the usual standard cosmology described by Einstein gravity.
However, we allow the value of the Planck mass to be
$\tilde m_{pl}\ne M_o$.
In this case, eqn \two\ becomes familiar,
$H^2+\kappa/R^2=(8\pi/3)\rho/ \tilde m_{pl}^2$.
For $C=0$, the curved cosmology is easy to understand:
if $\kappa=+1$ the universe is closed.  The expansion will eventually
cease and contraction will begin.  If $\kappa=-1 $
the cosmology is open.
The universe expands forever and is infinitely large.
If $\kappa=0$ the universe is flat. The expansion will slow
asymptotically to zero.
If $C\ne 0$, the description of the universe's evolution is more
complicated.  Still, we expect that adding some energy density
in a scalar field to the total energy density should not alter
the rough behavior of the universe with $\kappa$.
We verify that in fact the evolution of the Brans-Dicke universe for
the three values of $\kappa$ does correspond
to open, closed, and flat cosmologies in a familiar way.

Solving  the quadratic equation
\two \ for $H$ with $C\ne 0$ and $\kappa\ne 0$ gives
	\eqn\repewithk{	H=-
	{\dot \p\over 2\p}\pm\left [
        \sqrt{{(1+2\om/3)\over 4}\left ({\dot \p\over \p}\right )^2+
	 {8\pi\over 3\p}\rho-{\kappa\over R^2}}\ \right ]\  \  .
	}
Notice  that the $\pm$ here refers to the
two solutions of the quadratic
eqn \two \ for $H$.
We need to decipher which solution in eqn \repewithk\
corresponds to a growing solution; that is, a positive
Hubble expansion.

In the case of a flat universe, with
$\kappa=0$, the square root in eqn \repewithk \
is necessarily larger than the first term.
Thus, if we intend to study the expanding phase
( $H>0$ ), then we must choose  the solution
with the positive square root and so choose  the
$+$ sign.
Eqn \repewithk \ becomes
    \eqn\soldiers{
     H=-{\dot \p\over 2\p}+\left [
     \sqrt{{(1+2\om/3)\over 4}\left ({\dot \p\over \p}\right )^2+
	 {8\pi\over 3\p}\rho-{\kappa\over R^2}}\ \right ]\  \  . }
Since we are studying the radiation dominated era, we
use $\rho=\gamma T^4=\bar S^{4/3} \gamma/R^4$
where $\bar S$ is defined in eqn \sox .
Also, we pull a factor
of ${1\over 4} \left ({\dot \p\over \p}\right )^2(1+2\om/3)$
out of
the square root in eqn \soldiers \ to write
   \eqn\fus{      {H=-{\dot \Phi \over 2\Phi}\left
      [1+\ep{\sqrt{1
        +{\bar S^{4/3}\gamma\over \ep^2 R^4}
        \left({\Phi \over \dot \p^2}\right )
   -{\kappa \over \ep^2 R^2}
        \left({\Phi \over \dot \p }\right )^2
     }}\ \right ]
       }  \ \ ,}
where we define
	\eqn\ant{\ep=\pm {(1+2\om/3)\over 2}\ \ .}
 The $\pm$ in the definition of $\ep$ is
needed to ensure that $-(\dot \Phi/\Phi)\ep>0$ so that
only the growing solution
for $H$ with the positive square root is considered.
Thus the upper sign corresponds to ${\dot \p \over \p }<0$
and the lower sign corresponds to ${\dot \p \over \p }>0$.
There are therefore two distinct sets of $\pm$.
The first appears in eqn \repewithk\
and distinguishes the two solutions for $H$ which
solves the quadratic eqn \two .
The second set of $\pm$ in the definition of $\ep$
are needed to ensure that only the
solution for $H$ with  positive square
root is considered.

If $\kappa=-1$, $H$ is again positive only if the
$+$ sign is chosen in eqn  \repewithk .
However, if $\kappa=+1$,  it is possible that
the square-root is not larger than $\dot \p/2\p$.
If this is the case, then the negative square root can
yield  a positive Hubble expansion.
We will show in $\S$IIIC that the growing solutions with
negative square roots evolve from solutions
which at earlier times
obeyed eqn \repewithk \
with a positive square root.
This will be analyzed in detail when we study
the overall behavior of a $\kappa=+1$ cosmology.  In the
end we will find that
we can begin
with the positive square root in \repewithk\
for growing solutions.
We can proceed to solve for $R(\Phi)$ from eqn \fus.

We define the quantity $\chi$ as
       \eqn\into{ \chi(\Phi)={\bar S^{4/3}\gamma\over \ep^2 R^4}
        \left({\Phi \over \dot \p^2}\right )\ \ .}
Using the first integral of motion \again\ to eliminate
$\dot \Phi$, $\chi$ becomes
\eqn\repetwelve{
	       {\chi(\Phi) =  \bar S^{4/3} \gamma C^{-2}
        \ep^{-2}  \p R^2 \ \ }
         }
and we note that $\chi$ is always a real positive quantity.
We also define
	 \eqn\q{
      {Q^2={\ep^2 C^2\over \gamma^2\bar S^{8/3}}\kappa\ \ .}
      }
With this definition, $Q^2>0$ corresponds to $\kappa=+1$,
$Q^2<0$ corresponds to $\kappa=-1$, and
$Q^2=0$ corresponds to $\kappa=0$.
We rewrite \repewithk \ with these definitions
      \eqn\withk{
       {H=-{\dot \p\over 2\p}\left [1+ 2\ep
        \sqrt{1+\chi-Q^2\chi^2}\right ]\ \ }
      \ \ .
        }
Using $H=\dot R/R$, the definition of $\chi$, and rearranging,
we are left
with the integral
  \eqn\appendint{
       {\int_{\chi_i}^\chi
       {d\chi'\over \chi'\sqrt{1+\chi'-Q^2\chi'^2}}=-2\ep
        \int_{\Phi_i}^\Phi
        {d\Phi' \over \Phi'}\ \ .}
       }
Integrating this equation
we find
	   \eqn\usef{
      \chi={1\over \sinh^2\Theta  +Q^2\exp{(-2\Theta)}}
     \ \ ,
      }
where
we have absorbed the constants of integration
into the constant $\tilde \Phi$
	\eqn\errorr{
	\tilde \Phi=\Phi_i\left
	[{2+\chi_i+2\sqrt{1+\chi_i-Q^2\chi_i^2}\  \over
	\chi_i }\right]^{1\over 2\ep}
	}
and we define
    \eqn\wake{\Theta  =\ep\ln(\p/\tilde \p)\ \ .}
The relationship between $\Phi$ and $\tilde \Phi$ depends
on the value of $\kappa$.  For instance,
if $\kappa=0$ and the universe is flat,
then $\Phi$ asymptotically approaches the value
$\tilde \Phi$.
For $\kappa=\pm 1$ on the other hand $\tilde\Phi$ does not define an
asymptotic value.
For details see
\S IIIB and \S IIIC.

Using
$R=(\ep C/\bar S^{2/3})(\gamma\p)^{-1/2}\chi^{1/2}$, from the
definition
of $\chi$, we find
   \eqn\curv{
     {R={\ep C\over \bar S^{2/3}\gamma^{1/2}}
     {1\over \p^{1/2}}
        \left\{ {1 \over
     \sinh^2\Theta  +Q^2\exp{(-2\Theta)}}
      \right \}^{1/2}\ \ .}
      }
The temperature of the universe is found from adiabaticity to be
        \eqn\toe{
      {T={\bar S\gamma^{1/2}\over \ep C}\p^{1/2}
        \left\{ {
     \sinh^2\Theta  +Q^2\exp{(-2\Theta)}}
      \right \}^{1/2}\ \ .}
       }
The Hubble constant in terms of $\Theta$ is
     \eqn\pasta{
      H=\gamma^{1/2}{T^2\over \p^{1/2}}\left \{
      {\sinh^2\Theta+2\ep\sinh\Theta\cosh\Theta+Q^2(1-2\ep)
      \exp({-2\Theta})\over (
     \sinh^2\Theta+Q^2\exp({-2\Theta}))^{1/2}  }
			\right \}\ \ .
      }

Armed with these results we can now
discuss the nature of the solutions to the equations of motion
for the different values of $\kappa$.

\centerline{\bf IIIA.  A flat cosmology, $\kappa=0$}

In Ref [4] this example was worked out in detail.
We provide only a brief description here for completeness.
There are three possible initial conditions for $m_{pl}$.
The Planck mass could have the constant value denoted
$\tilde m_{pl}$
throughout the
radiation dominated era.
Alternatively, $m_{pl}$ could
start  out
initially small
and grow.
Lastly, $m_{pl}$ could  initially be large
and drop.
In both of these later cases $m_{pl}$ approaches the
boundary value $\tilde
m_{pl}$ as the scale factor grows.  As can be seen from
eqn \again ,
as the scale factor grows infinitely large,
$\dot \p\rightarrow 0$ and
the change in the Planck mass shuts off.

This general behavior is illustrated in figure 1 which shows
schematically $R$ as a function of $\Theta$.
Notice that time is increasing along the horizontal axis
from right to left.
As the Planck mass approaches the asymptotic value
$\tilde m_{pl}$ and thus $\Theta$ decreases
toward  zero,  the scale factor grows.

 While $\dot
\p$ is significant, the scale factor and the temperature evolve with the
changing $m_{pl}$ in a complicated way.
Once $m_{pl}$ veers close to its asymptotic value $\tilde m_{pl}$, then
$d m_{pl}/dt\approx 0$ and the universe evolves in a familiar way.
For
$m_{pl}\approx \tilde m_{pl}$ roughly constant,
the equations of motion
reduce to those of an ordinary radiation dominated Einstein
cosmology with
$M_o$, the usual Planck mass of $10^{19}$ GeV, replaced with $\tilde
m_{pl}$. In particular, this means $R\propto t^{1/2}$, $T\propto t^{-1/2}$,
and $H=1/2t$. Despite the underlying structure of the  theory, gravity
appears to be described by a standard
flat universe with a static gravitational constant.
The universe will expand  forever, slowing
with age to almost a halt.

\centerline{\bf  IIIB. An open cosmology, $\kappa=-1$}

If  $\kappa=-1$
and so $Q^2<0$,  the cosmology is open in the usual sense
as it expands forever.
This is confirmed by the expression
for $R(\Theta)$ .
In figure 2 we plot $R$ of eqn
\curv\ as a function of $\Theta$.
Time increases from right to left
along the horizontal axis.
According to our calculation, $\Theta$ is always positive
and decreasing.
Initially $R=0$ at $\Theta=\infty$.
As $\Theta$ drops $R$ increases.

Let $\Theta_M$
be the minimum value of $\Theta$.  At $\Theta=\Theta_M$, the denominator
in \curv \ vanishes and $R\rightarrow \infty$.
According to the first integral of motion \again ,
$\dot \p\rightarrow 0$ when $R\rightarrow \infty$
and the change in $\p$,
and thus also in $\Theta(\Phi)$,
turns off.
The denominator in \curv \
vanishes at
	    \eqn\vanquish{
      {\sinh^2\Theta_M+Q^2\exp{(-2\Theta_M)}=0}
      }
which gives
 	\eqn\oiplease{
       {\Theta_M={1\over 2}\ln\left [{1 + \sqrt{-4Q^2}}\right ]
       \ \ .}
        }
For $\Theta\ge \Theta_M$, $R$ is real.
Notice that since $\Theta_M>0$, we know that the Planck
mass never reaches the value $\tilde m_{pl}=\tilde \Phi^{1/2}$.

A rough sketch of the history
of an open Brans-Dicke cosmology begins
with
 $\Theta=\infty$ and
$R(\Theta)=0$.
As $\Theta$ drops  toward $\Theta_M$ $R$ grows.
The universe expands forever, growing infinitely large as $\Theta$
approaches $\Theta_M$.
There is no possibility for the
Hubble expansion to vanish.
Thus the gross behavior of
this cosmology is similar to  that of a standard open ($\kappa=-1$)
cosmology.

\centerline{\bf  IIIC.  A closed cosmology, $\kappa=+1$}

In the standard
closed
model, $\kappa=+1$,
the universe will grow to
a maximum
size, thus cooling to the minimum temperature
$T_{\rm col}$ defined later in
 eqn (40), at which point the
expansion ceases and contraction begins.
In a Brans-Dicke model,
for $\kappa=+1$
($Q^2>0$), the cosmology is also
closed.
As in the standard
model the expanding phase eventually ends as the contracting
phase begins.
If $\dot \p/\p=0$, then
the Planck mass is constant and it is simple to see that
the standard behavior is reproduced with $M_o$ replaced
by  $\tilde m_{pl}$.  If $\dot\p/\p\ne 0$, then
the closed
cosmology is a bit more subtle than in the standard model
and it takes a bit of work to see this general
behavior.

Firstly,
solve \two\
for $H$ to rewrite eqn \repewithk \  with the abbreviations
$H_R^2\equiv (8\pi/3\p)\rho$ and $\mu\equiv \dot \p/\p$
				\eqn\reex{
				H=-\mu/2\pm\sqrt{\ep^2\mu^2
						+H_R^2-\kappa/R^2}\ \  ,					}
where again the $\pm$ here refers to the two solutions of the
equation of motion quadratic in $H$.
If $H$ is to reach zero at a finite temperature
and reverse sign so the universe collapses,
then it is critical that $-\mu/2$ and the
square root have opposite signs.
It seems conceivable that, for instance, both
$-\mu/2$ and the square root will have the same
sign and $H$ will never vanish.  We will find in
the end that all is well; $H$ will in fact reach zero and reverse course,
but some effort will be required to demonstrate this fact.
We will establish in the next subsections that the
solution to  eqn \reex\ with negative square root
evolves from the growing solution to eqn \reex\
with the positive square root.
We study the
two possibilities, $\mu>0$ and $\mu<0$, separately.

For future reference, we write here
the most pertinent results which
will be derived below.
In the end, it will be shown that the universe
does stop expanding for $\kappa=+1$
and begins to collapse, regardless of $\mu$,
at a temperature of
       \eqn\helpreview{
       {T_{\rm col}={\p_{\rm col}^{1/2}\over
      \gamma^{1/2}\bar S^{1/3}} {1\over \left |Q\chi_{\rm col}^{1/2}\right
         |}\ \  ,}
        }
where
       \eqn\woopreview{
         {\chi_{\rm col}={1\over 2Q^2}
         \left [ 1+\sqrt{1+Q^2{(4\ep^2-1)\over
         \ep^2}
         }\right ]\ \ }\ \ ,
           }
and
         \eqn\helper{
        {{\p_{\rm col}
       \over \tilde \p}=\left [{(2+\chi_{\rm col})\ep+1\over \chi_{\rm col}
       \ep}
        \right ]^{1\over 2\ep}\ \ .}
          }

\centerline{\bf  IIIC.1. Closed Brans-Dicke cosmology
with $\bf \dot \Phi/\Phi>0$}

We show in this subsection that if $\kappa=+1$, the Hubble
expansion will eventually end and the universe will ultimately
collapse
for the case of $\mu=\dot \Phi/\Phi>0$.
In the next subsection we repeat the analysis
and verify the same evolution is predicted for the case of  $\mu<0$.
As well,
we derive the results \helpreview - \helper\ here.

In this subsection we take $\mu=\dot \p/\p>0$ so that
$\Phi$ grows with time.
Consider the evolution of the three terms under the
radical in eqn \reex :  the kinetic term $\ep^2 \mu^2$,
the radiation term
$H_R^2$, and the curvature term $-\kappa/R^2$.
 From the first integral of motion in the
eqn \again , we can see that the kinetic term
scales
as $\ep^2 \mu^2\propto
R^{-6} \p^{-2}$.  From its definition we know that $H_R^2$
scales  as
 $H_R^2\propto R^{-4}\p^{-1}$
while the curvature term $\propto 1/R^2$.
As we look back in time,
$\p$ gets smaller  with $R$.  So,
tracing back to $R\rightarrow 0$
for the sake of argument, we see that the kinetic
term  dominates over both the other terms initially and drops the most
quickly.  The next dominant term is $H_R^2$
which drops more slowly than the kinetic term but more quickly
than the curvature.  The curvature term is the least important
of the three initially.  Eventually, as $R$ grows curvature
gains importance.

 If $\p$
is growing  then $H>0$
only if we choose the positive square root in \reex .
With this choice of signs, eqn \reex \ becomes
$H=-|\mu/2|+\sqrt{\ep^2\mu^2+H_R^2-{\kappa/R^2}}$.
In the beginning,
 when the scale
factor is quite small, the curvature term is much
less important than the
sum of the positive terms
in the square-root.  This must be so for
the square-root  to exceed $|\mu/2|$
and thus lead to an expanding universe, at least initially.
Note that $H$ can vanish and will eventually do so.
$H$ will vanish and the expansion cease when the
square root equals $\mu/2$.

In figure 3 we have a schematic picture  of the
development
of the sum of positive terms versus the development
of the absolute value of the curvature term, $1/R^2$.
For $\mu>0$, the case we study here, the universe starts to collapse
(i.e. $H=0$) at the point indicated on the figure.
Collapse begins before
the sum of positive terms crosses the  curvature term, i.e.,
before the square root vanishes.

The value of $T$
at which $H$ reaches zero
can be found by setting
 $H=0$ in \reex \
and solving for the temperature.
Remember
in eqn \toe \  the temperature is expressed
completely in terms of the value of
$\Phi$, up to the constants $\ep, C, S$ etc.
Instead of referring to the collapse
temperature, we could equally well refer
to the value of $\Phi$ at which $H=0$, $\Phi_{\rm col}$.
To find $\Phi_{\rm col}$ we first
set $H=0$ in  \withk \ and solve for the the maximum value of
$\chi$ (see eqn \repetwelve ), called
$\chi_{\rm col}$,
       \eqn\woo{
         {\chi_{\rm col}={1\over 2Q^2}
         \left [ 1+\sqrt{1+Q^2{(4\ep^2-1)\over
         \ep^2}
         }\right ]\ \ }\ \ .
           }
This can then be used in the definition of the temperature in \toe \
to find the temperature at which the universe begins to collapse,
       \eqn\help{
       {T_{\rm col}={\p_{\rm col}^{1/2}\over
      \gamma^{1/2}\bar S^{1/3}} {1\over \left |Q\chi_{\rm col}^{1/2}\right
         |}\ \  .}
        }

 From expression
\repetwelve \   for $\chi$ and the definition of
$\Theta$, we see this corresponds to a
maximum value of $\p$ for $\mu >0$
         \eqn\helpagain{
        {{\p_{\rm col}
       \over \tilde \p}=\left [{(2+\chi_{\rm col})\ep+1\over \chi_{\rm col}
       \ep}
        \right ]^{1\over 2\ep}\ \ .}
          }
Recall that $\ep$ was defined in eqn \ant\ so that the product
$-\mu \ep>0$.
For the case of $\mu>0$ treated here, then
$\ep<0$ and
\helpagain \ is less than 1.
In terms of $\Theta\equiv\ep\ln(\p/\tilde \p)$,
eqn \helpagain \ implies $\Theta_{\rm col}\ge 0$.

Once $\p$ reaches $\Phi_{\rm col}$,  which is $\le \tilde \p$,
then the expansion ceases and the universe begins collapse.
Notice from the first integral of motion
\again , that $\Phi$ continues
growing beyond $\Phi_{\rm col}$ as the universe contracts.
In plot 4 we show the rough behavior of $R$ with $\Theta$
where again time increases from right to left.
The scale factor hits a maximum at $\Theta_{\rm col}\ge 0$
and begins contracting.
Notice as $\Phi$ continues to grow, $\Phi$ can exceed the
value $\tilde \Phi$ in the definition of $\Theta$ and
thus $\Theta$ can
become negative.

We have not yet shown that the collapse temperature
and collapse $\p$ defined here have relevance for $\mu<0$
but we do so in the next section.
For later reference we notice that if $\mu<0$
and  $\ep>0$, then  \helpagain \
is greater than 1 and again $\Theta_{\rm col}\ge 0$.
An analogous picture to plot 4 applies for the closed
universe with $\mu<0$ discussed next.

\centerline{\bf  IIIC.2. Closed Brans-Dicke cosmology with
$\dot\p/\p<0$}

If $\mu<0$ and
$\p$  is  dropping then the analysis is a bit
more complicated but the end result is very similar.
We start with the assumption that initially the positive square root
in \reex \
gives
a real expanding cosmology and show that this is self-consistent.
Consider again the three terms
under the square-root of
\reex :
$\ep^2\mu^2
\propto R^{-6} \p^{-2}$,
$H_R^2\propto R^{-4}\p^{-1}$, and $\kappa/R^2$.
Since $R$ grows while $\Phi$ drops, there is a competition in the
denominator of the kinetic term.  There is a similar
competition in the denominator of the
radiation term.
We will show here that in fact $\Phi R^2$ grows when the
square root is positive
and therefore establish
that kinetic and radiation terms drop as the universe evolves.
To get a handle
on this notice that the equation of motion
\reex , with the positive square root, can be rearranged to read
     \eqn\oncemo{
     H+{\mu\over 2}={\dot R\over R}+{\dot \Phi\over 2\Phi}
    ={1\over 2}{d\ln(\Phi R^2)\over dt}=+\sqrt{\ep^2\mu^2
						+H_R^2-{\kappa\over R^2}}
	\ \   .
		  }
This shows explicitly that $\Phi R^2$
grows with time.
Looking back in time, $\Phi R^2$ drops and so
$\ep^2\mu^2
\propto R^{-6} \p^{-2}$ grows as we go back in time.
We also note that $\ep^2\mu^2\propto
H_R^2 (1/\Phi R^2)$.
We can conclude then that
if we trace back to $R\rightarrow 0$ that initially
 the kinetic term
$\ep^2\mu^2$ dominates over
$H_R^2$ for very small
values of $R$
and loses its importance as $\Phi R^2$ grows.
Notice $H_R^2\propto \kappa/R^2(1/\Phi R^2)$
and so, by pursuing the same reasoning,
we see that in turn  $H_R^2$
dominates over the curvature.
Again, the curvature term is the least important
of the three initially.
We then begin with the positive square root in \reex .

At first glance it seems
that $H$ will not go to zero
at finite temperature, $H=|\mu/2|+\sqrt{\ep^2\mu^2
	+H_R^2-\kappa/R^2}$.
However,  as $R$ grows, curvature
eventually
gains importance
and the
square-root passes through zero.
[Eqn \oncemo \ shows that
$\ln(\Phi R^2)$
has  an extremum
when   the square root vanishes.
Taking the second derivative of
$\ln(\Phi R^2)$, evaluated when
the square root vanishes, we see that
the extremum is a maximum of $\ln(\Phi R^2)$.
In other words,  the first derivative of $\ln(\Phi R^2)$ passes through
zero and then becomes negative.  We can make the connection that the
square root in eqn \oncemo\ is equivalent to the first
derivative of $\ln(\Phi R^2)$ and so we know that
the square root
falls smoothly through
zero, becoming negative.]
The solution for $H$ is then eqn \repewithk \ with the
negative square root,
$H=|\mu/2|-\sqrt{\ep^2\mu^2
	+H_R^2-\kappa/R^2}$.
We find that solutions to $H$ with negative square root grow out of
solutions to $H$ which began with positive square root.

As the magnitude of the square-root grows it eventually
balances the $\mu/2$ term until $H=0$.
The expanding phase ends and the
universe begins to contract.
This will happen at the same
collapse temperature as defined in
eqn  \help \
for $\mu>0$.
Thus $T_{\rm col}$ of eqns \helpreview \ and \help \
is the general expression defining the temperature
at which a closed  Brans-Dicke universe begins to contract.
As $\Phi$ drops to the value $\p_{\rm col}$, which is
 $\ge \tilde \p$, the expansion ceases and reverses direction.
As the universe collapses $\p$ continues to drop.

Before we close this section, we note that we traced back
to $R\rightarrow 0$ to draw conclusions.
We cannot actually trace back all the way to $R\rightarrow 0$ since
we would enter the epoch of quantum gravity at some finite $R$.
If instead we start the evolution of the universe at finite $R$ then the
relative importance of the terms contributing
to the square root depends on the relative amplitudes.
For $\mu<0$, in principle we could begin
 at finite $R$ with positive solution for $H$ with a negative
square root.
What we have shown is that in general
solutions to $H$ with negative
square root grow out of solutions to $H$ which
began with the positive square root.
In figure 5
is drawn a schematic picture of the sum of the positive terms versus
the magnitude of the curvature term.
Along the horizontal axis time grows from left to right.
This figure shows that the universe collapses after  the
sum of positive terms equals the curvature term
and the square root
vanishes, as we have argued above.
 To the left of the crossing
point, the square root is positive while to the right
of the crossing point, the
square root is negative.
In principle, for $\mu<0$, as this figure shows,
one could begin with the universe at finite $R$
between the points in figure 5 when the square root
vanishes and the universe begins to collapse.
So one could begin with growing solutions ($H>0$)
with negative square root, for some range of parameters.

Now that we have a picture of the large-scale behavior
of a curved cosmology in a theory of modified gravity, we can
discuss the flatness problem in these theories.
We will work in analogy with the standard model and
so build the framework for the standard flatness discussion here.

\centerline{\bf  IV. The flatness problem in the standard model}

We argued in
the introduction that,
generically, adiabatic cosmologies will have to contend with
a large $\bar S$ and so a flatness problem.
In this
section we interpret flatness for the standard cosmology
in terms of the
early cosmic dynamics and the energy density of the universe.$^{12}$

Consider a closed cosmology
so that $\kappa=+1$.
According to the standard Einstein equations, $H^2+{\kappa\over R^2}
={8\pi\rho\over 3 M_o^2}$, for $\kappa=+1$
the expansion ceases and
the universe then starts to collapse at a temperature of
	\eqn\stop{
	{T_{\rm col}={M_o\over \gamma ^{1/2}
	\bar
	S^{1/3} }\ \ }
		}
with $\gamma\equiv(8\pi^3/90)g_*(t)$
and $g_*(t)$ is the number of
relativistic degrees of freedom in equilibrium at time $t$.
For ease of notation we again use the definition
$ \bar S=R^3T^3 $ where $S \simeq \bar S
(4/3)(\pi^2/30)g_*$, and $S$ is the constant entropy.
For moderate values of $ \bar S$,
then $T_{\rm col}\sim M_o$.
 At a temperature of $\sim M_o$ the universe would
reach its maximum extent and
then implode.
If we require that the
universe continues to expand
until today so that $T_{\rm col}<T_o$, then
it must be that
$\bar S^{1/3}\gta {M_o/T_o}\sim 10^{32}$.
Thus the arbitrary constant entropy of the standard big bang
model must be extraordinarly large if the universe is to
survive until
a temperature of $T_o\sim 2.74^o$ K .

We can relate the flatness
problem to the commonly used parameter
$\Omega\equiv \rho/\rho_{cr}$,
$\rho$ is the total energy density of the universe,
and $\rho_{cr}$ is the critical value required to just close the
universe; that is,
$\rho_{cr}$ is that value of the energy density required to just balance
$H^2$ if $\kappa=0$,
  	\eqn\criticalv{
	{8\pi\over 3M_o^2}\rho_{cr}=H^2 \ \ .
	}
Numerically,
$\rho_{cr} = 1.88
\times 10^{-29} h_o^2$ gm cm$^{-3}$, where
$h_o = H_o / 100$ km s$^{-1}$
Mpc$^{-1}.$

 According to the standard
Einstein equations, we can write for general $\kappa$
       \eqn\youzayy{{  \Omega-1={\kappa\over H^2R^2}\ \  .}}
If $\kappa/H^2R^2\rightarrow 0$
and the cosmology is nearly flat,
then  $  \Omega\rightarrow 1$.
Written another way,
	\eqn\anty{
	{  \Omega={1\over 1-x(t)}\ \ }
	}
and
	\eqn\reduic{
	{x={M_o^2 \kappa\over \gamma \bar S^{2/3}T^2}\ \
	\ {\rm or}\ \ \ x=\left ({T_{\rm col}\over T}\right )^2 \kappa
	\ \ .}
	}
For the closed cosmology of $\kappa=+1$, eqns \anty \ and
\reduic \ say that
when $T=T_{\rm col}$,
$x=1$ and $\Omega\rightarrow
\infty$.
Thus $\Omega\sim1$ is unstable and
$\Omega$
will
quickly diverge for temperatures below $T_{\rm col}$.
In terms of $\Omega$,
a large value for $\bar S$ means
a small collapse temperature and so a
small $x$.
A small $x$ in turn renders $\Omega\sim 1$,
 corresponding to
a nearly flat universe.

So the flatness problem can be stated
in terms of $\Omega$.
As $\Omega\sim 1$
is very
unstable, it is unlikely and in some
sense unnatural
for it to be near 1 today.
The observations that today $\Omega_o \sim 1$ would require,
for instance,
at a temperature of the Planck scale,
that $\Omega (T=M_o) - 1 \simeq
O(10^{-60})$.  In words,
for $\Omega$
to be of order 1 today requires the universe to be
created with the extreme condition that initially
$\Omega$
 be identical
to 1 to better than one part in $10^{60}$.

Similarly, in a standard open cosmology for which $\kappa=-1$
there is a
flatness problem. Near a temperature of $T_{\rm col}$ given in
\stop , $x(T_{\rm col})\sim-1$ and $  \Omega\sim 1/2$
which, astrophysically speaking, is on the order of 1.
For temperatures $T<T_{\rm col}$,
the universe will not collapse as in the closed
case.
However,
$x$ gets large and negative
as the temperature drops below $T_{\rm col}$
and
this drives $\Omega\rightarrow 0$.
Thus, even for a
standard open cosmology, the temperature
defined as $T_{\rm col}$ represents the temperature
at which $\Omega\sim 1$ becomes unstable.
 If today
$ \Omega_o\sim 1$  then  today
$0\ge x(T_o)\gta-1$.
The requirement that  $  \Omega_o\sim 1$ today
demands that $T_{\rm col}<T_o$ which in turn demands that
$\bar S^{1/3}\gta M_o/T_o$.
If this were not the case, the universe would cool
to the low temperatures of today in a Planck time, i.e.,
$10^{-43}$ sec.

\centerline{\bf  V. Defining $\bf \bar \Omega$ for scalar gravity}

In $\S$ IIIC, the collapse temperature was defined
in eqn \helpreview\  for a closed
Brans-Dicke cosmology.
Before addressing the flatness problem in the
Brans-Dicke model, we first develop
the last tool needed and define here a new
measure of the energy density of the universe, $\bar \Omega$.

We want to cast a
flatness argument
in analogy with the
treatment for the standard cosmology.
To do so, we here
define a quantity $\bar\Omega\equiv\rho_{tot}/\rho_{cr}$
where $\rho_{tot}$ is the sum of all energy densities,
including the energy density in $\dot \p$,
and
where
$(8\pi/3\p)\rho_{\rm cr}=H^2$ corresponds to the value of the total
energy density required to
just close the universe.
Equivalently,  $\bar \Omega\equiv (H^2+\kappa/R^2)/H^2$
or using eqn \two ,
	\eqn\nomore{
	{\bar \Omega={{8\pi\rho\over 3\p}-{\dot\p\over \p}H+
	{\om \over 6}
	\left ({\dot\p\over \p}\right )^2\over {8\pi
	\rho_{\rm cr}\over 3\p}}\ 		\  .}
	}
With this definition for $\bar \Omega$ we can
write the equation of motion \two \ as
	\eqn\youza{
	{\bar \Omega-1={\kappa\over H^2R^2}\ \ .}}
If
$\kappa/H^2R^2\rightarrow 0$
then
 $\bar \Omega\rightarrow 1$.
Written another way,
	\eqn\another{
	{\bar \Omega={1\over 1-x(t)}\ \ }
	}
and
	\eqn\ho{
	{x={\kappa/R^2\over {8\pi \rho\over 3\p}-
	{\dot\p\over\p}H+{\om \over 6}\left ({\dot\p\over
	\p}\right )^2}
	={\kappa/R^2\over H^2+\kappa/R^2}\ \ .}
	}
With some work we can rewrite
$x$  as
	\eqn\stupid{{x=
  	{\kappa\over
  	\left (T/T_{\rm col}\right )^2\left [1+
	\left (T^2/\p\right )\chi_{\rm col}^{-1}
	\left \{ 1+2\ep\sqrt{1+\chi-Q^2\chi^2}\right \}\right ] }\
	\ \ .}
	}
There are several things to notice about these expressions for
$x$.  Firstly,
for $\kappa=+1$,
the far right hand side of eqn
\ho\ makes clear that
 $x=1$ at $H=0$ and so
$x=1$ when $T=T_{\rm col}$.
At
$x=1$, $\bar \Omega\rightarrow
\infty$
according to \another .
This adheres to our expectations.
At the collapse temperature $\bar \Omega\sim 1$ becomes
unstable.

In an open cosmology ($\kappa=-1$), the universe
does not collapse as it does in the closed
($\kappa=+1$) universe.  Still,
near a temperature of $T_{\rm col}$ given in \help ,
$x(T_{\rm col})\sim-1$ and $\bar \Omega\sim 1/2$.
For temperatures $T\ll T_{\rm col}$,
$x\rightarrow -\infty$ and $\Omega\rightarrow 0$.  If
$\bar \Omega_o\sim {\cal O}(1)$ today then
$0\ge x(T_o)\gta{\cal O}(-1)$ today.
Thus, the requirement that  $\bar \Omega_o\sim 1$ today
brings the same conclusion to that of the $\kappa=+1$ case.

\centerline{\bf
VI.  The flatness problem in  Brans-Dicke cosmology}

The flatness problem in a  Brans-Dicke cosmology
can be quite complicated.  Here we take
$m_{pl}\approx
\tilde m_{pl}$ to move slowly
and to be near the value $M_o=10^{19}$ GeV, so that there
is little deviation from standard Einstein gravity.
These assumptions greatly simplify the discussion.
We will discuss in the next sections a Planck mass
far from the value $M_o$.

As for the standard model,
we discuss the huge entropy condition
in terms of a collapse temperature and consider first
the closed cosmology ($\kappa=+1$).
With $\dot \Phi/\Phi\approx 0$ the universe evolves as in
a standard cosmology with $M_o$ replaced by $\tilde m_{pl}$.
If we study our example of the closed cosmology again, we
find the collapse temperature of eqn \stop \
reduces to
	\eqn\collos{
	{T_{\rm col}={\tilde m_{pl} \over \gamma ^{1/2}
	\bar
	S^{1/3} }\ \ .}
	}
For moderate values of $\bar S$, the universe would contract
at a temperature near the Planck scale.

Although eqn \collos\ only holds true
during radiation domination we can easily
correct for the era of matter domination
to have a rough indication
of the condition through to today.
If we want the universe to survive until today,
then $T_{\rm col}\lta T_o$  with $\tilde m_{pl}\sim
M_o$ and it must be that
$\bar  S  \gta 10^{90}$.
Of course, we should guess that the standard model
flatness problem surfaces here since we
assumed the cosmology look like standard Einstein gravity,
A more complicated situation arises if we
allow for a large deviation from Einstein gravity.  In particular,
a conflict arises for \mad gravity which  tries to exploit
modified
gravity to address the horizon and monopole problems.
Although the universe is in principle made flatter in \mad gravity,
the flatness problem is not solved.
We delve into this subject next.

\centerline{\bf  VII.
MAD gravity and the Horizon, Monopole, and Flatness Problems}

The standard
cosmology does not explain the remarkable
smoothness and flatness of the observed universe.
We can presently see across many regions which were not in causal
contact at earlier times.
All the same, today the universe does
seem to be largely homogeneous and isotropic.
This apparent smoothness
seems to violate causality.
As well, the universe
appears to be roughly
flat today;
that is,
matter continues to be important
in determining the cosmic evolution so it must be
that curvature does not completely dominate.
In the standard cosmology, a
universe which began with arbitrary initial conditions
would quickly veer away from a flat appearance.
In the absence of a dynamical explanation,
a nearly flat universe
today requires extraordinary initial
conditions
which render  the universe extremely close to flat at early times.
In addition,
the inclusion of grand unified theories
into the standard cosmology
gives rise to a
cosmologically disastrous abundance of monopoles.
Although the monopole problem has a very
different source from the horizon problem and flatness problem,
solutions to one  are often intimately connected with
solutions to the others.

The inflationary model
proposed by Guth addresses the horizon, flatness, and monopole
problems.
In
the inflationary scenario, a potential energy density drives a period
of accelerated growth of the scale factor.
During this period,
a causally connected region that was
small at the beginning of inflation grows
large enough to contain our observed universe.
Then the homogeneity of the observed universe can be explained by a
common history. As the universe inflates, the monopole abundance  is
diluted, as is everything else.
Subsequent to this era of supercooling, entropy is produced
as the potential energy is converted
to radiation and the universe resumes an ordinary evolution.
The generous
entropy production reheats the universe to some high temperature,
arranged to be below the temperature at which monopoles form.
Thus inflation explains the present homogeneity and lack of monopoles.
In addition,
an inflationary epoch also
allows the universe to begin with
moderate initial values for the entropy.
The enormous value of the entropy needed to explain the
cosmic flatness today is generated dynamically.

In Ref [4] and [5] we
suggested that a cosmology with a dynamical Planck
mass, such as the Brans-Dicke model studied in this paper, can provide
an alternative resolution to the horizon and monopole
problems, though not the flatness problem.
The horizon problem is resolved by slowing the evolution
of the universe
during the era of radiation domination.
Early in the universe's history the structure
of gravity slows the Hubble expansion, thus
slowing the cosmological evolution.
As a result, the universe at a given temperature
is much older than in the standard
model.
Thus
enough
time elapses for the entire observable universe
to be in causal contact.
Large regions could thereby become smooth
without violating causality.
Expanding the horizon can also dilute the monopole
density. As well, the slow Hubble expansion
keeps monopole-antimonopole annihilations in equilibrium
longer, allowing for a very low relic monopole abundance at
the end of the day.

Adiabaticity was assumed in the original formulations
of MAD gravity to make clear the role of the dynamical
Planck mass.
In Ref [8] and [9] obstacles to completing
the adiabatic \mad picture are discussed.
Some of these obstacles could be circumvented if the assumption of
adiabaticity  is removed
or if higher order theories of gravity are
considered.$^{10}$
Regardless of the troubles the \mad model faces,
it is always the case that adiabatic \mad gravity will not
address the flatness problem.
The persistence of a
flatness problem in the
Brans-Dicke model is a direct consequence of the
assumption of adiabaticity.

Since the flatness problem and the horizon problem are related,
we first introduce the horizon problem
and sketch the \mad prescription.
A causal explanation of the homogenity of our observable
universe could exist if a
region causally connected at some high temperature
grows big enough
to encompass everything we can see.
Since we can see back to the time
of decoupling, the size of the observable universe
is roughly the distance light could have
traveled since that time, $\Delta t_o \sim H_o^{-1}$,
where $H_o$ is the Hubble constant today.
Thus we can take the present comoving Hubble radius,
$1/(H_o R_o)$, as a measure of the
comoving radius of the observable universe.
The particle horizon defines the extent of a causally
connected region.
In the standard model the horizon $\sim H^{-1}$ so that
the causality condition
can be written as
  \eqn\horsta{  {1 \over H_cR_c}>{1\over H_oR_o}\ \ . }
The subscript $c$
denotes values at an early time
and subscript $o$ denotes values today.
[This equation only holds if  the horizon size, $d_{\rm horiz}$,
obeys $d_{\rm horiz}\sim H^{-1}$.
More generally the causality condition is
$d_{\rm horiz}(t_c)R_c^{-1}\gta d_{\rm horiz}(t_o)R_o^{-1}$.]
The observable
universe today fits inside a region
causally
connected at time
$t_c$ if
eqn \horsta\ is satisfied.
Then the horizon size at $t_c$ before
nucleosynthesis is large enough
to allow for a causal explanation for the
smoothness of the universe today.
Since $H=\dot R/R$, eqn
\horsta\ is equivalent to the requirement the $\dot R_o\gta
\dot R_c$; that is, the scale factor grows faster today than
at earlier times and thus there must have been a period
of acceleration between $t_c$ and today.

For Brans-Dicke gravity with general curvature,
the causality condition \horsta\ would require
      \eqn\causing{
       {{\p_c^{1/2}\over T_c}2\ep\left [{
      (\sinh^2\Theta_c+Q^2\exp({-2\Theta_c}))^{1/2}\over
      \sinh^2\Theta_c+2\ep\sinh\Theta_c\cosh\Theta_c+Q^2(1-2\ep)
      \exp({-2\Theta_c})
      }\right ]\gta \beta {M_o\over T_o}\ \ .}
     }
Notice that as $Q^2\rightarrow 0$ \curv ,
\toe ,
and \causing \ reduce to the corresponding results
for a flat
universe, as it must.  Similarly, for large $\Theta$,
$e^{-2\Theta}\rightarrow 0$, and
we have the
same causality condition as in the case of the
flat universe.

\centerline{\bf VIIB.  The \mad Slow Roll Limit}

In comparison to the complicated constraint
eqn \causing , consider the simplifying
assumptions  of a slowly rolling Planck
mass and a flat cosmology.  In the slow roll limit the condition \causing\
becomes much more simply
    \eqn\hor{
     {m_{pl}(t_c)\over M_o} \geq \beta
      	{T_c \over T_o} . }
If the Planck mass were this large during an early hot epoch and
thus the strength of gravity was weak, then a  causally
connected region would have time to grow large enough to encompass
everything we can see.
Subsequent to $t_c$, the strength of gravity must grow as the
Planck mass drops.
In the absence of all entropy production, it is difficult
to drive the Planck mass from the large
value indicated in eqn \hor\	     to the value
$M_o$.$^{10, 11}$

In principle,
the disparity between the large early value of the
Planck mass needed to resolve the horizon problem
in the slow roll limit
and the Planck mass
today leads to a flatter universe.
As the Planck mass drops
after time $t_c$ and the
strength of
$G$ increases, the universe becomes flatter;
that is, since $G$ describes the strength with which matter affects
the cosmic development,
curvature becomes less important
than matter
as the coupling strength increases.
Still, the flatness problem
is not removed
entirely in a \mad era. Instead it is pushed to a higher energy
scale.
We study this question in detail here for a MAD Brans-Dicke theory.

First notice that
in terms of $\bar \Omega=1/(1-x)$
for $\mu$ identically zero, $x$ reduces to
	\eqn\reduc{
	{x={\p \over \gamma \bar S^{2/3}T^2}\kappa
	\ \
	\ {\rm or}\ \ \ x=\left ({T_{\rm col}\over T}\right )^2
	\kappa\ \  .}
	}
Since $\mu=\dot \Phi /\Phi$, the above expression can be taken
as an approximation in  the slow roll limit.
Between $t_c$ and today, $x$ changes by
      	\eqn\oiveyismere{
	{x_o\over x_c}\sim \left ({M_o\over  m_{pl}(t_c)}
      	\right )^2\left ({T_c \over T_o}\right )^2
      	\lta 1
	}
where the second relation follows from the horizon condition
for a nearly constant Planck mass, eqn  \hor .
In other words,
once the Planck mass reaches the value $M_o$, there is a
new collapse temperature,
$T_{\rm col}(M_o)=(M_o/m_{pl}(t_c))^2T_{\rm col}(t_c)$.
In the standard model on the other hand
$x$ would have grown by a factor
of $(T_c/T_o)^2$.
For example, if  $T_c = 10^{16}$ GeV,
$x$ would have grown by a monstrous factor of
$10^{55}$.
Thus, \mad assists the approach to flatness.

Although the universe gets flatter, there is still a flatness problem.
Consider a closed cosmology.
If the universe does not survive until the temperature
drops to $T_c$, then the \mad model does not
have the opportunity to address even the horizon problem.
We will therefore always require that the
universe continues to expand
until a temperature below $T_c$.
By the way, it is also true in an  inflationary
cosmology that the temperature
at which the universe begins to collapse must also be below the temperature
at which inflation begins.

In the slow roll limit, the collapse temperature is given
roughly by
       \eqn\rugg{T_{\rm col}={ m_{pl}\over \gamma^{1/2}
   \bar S^{1/2}  }\ \ .}
For the slow roll \mad model, $m_{pl}$ is many orders
of magnitude larger than in the standard model.
As a result of the huge Planck scale the temperature
at which the universe begins to collapse is correspondingly larger.
Given $T_{\rm col}<T_c$, eqn \rugg\
can be expressed as a condition on the entropy
 		\eqn\horses{\bar S^{1/3}\gta
{ m_{pl}(t_c)\over \gamma^{1/2} T_c}
			\ \ .}
The constraint on the Planck mass in the slow
roll limit for a \mad model which
addresses the horizon problem
in eqn \hor\ can be used to fix the
constraint on the entropy.
We find
      \eqn\sushi{
     \bar S^{1/3}\gta
	\beta     {M_o\over T_o}\ \ ; }
that is to say, $\bar S\gta 10^{90}$.
A large entropy is needed if the curvature of
the universe is not to take over just
below the very large Planck scale.
Although the universe gets flatter,
the initial requirement
of eqn \sushi\ that $\bar S^{1/3}\gta \beta M_o/T_o$ is not
alleviated. For $\bar S\sim 1$,
the huge Planck scale and thus early Planck time leads to the
instability of  $\bar \Omega\sim 1$ well above $T_c$.
Thus there is a flatness problem.

In an inflationary model where inflation
begins at $T_c = M_o$, the flatness problem
is solved.  On the other hand if $T_c < M_o$,
then an inflationary model may require
$S\gg 1$
in order for the universe to be able
reach the temperature at which inflation begins.
In particular, in a closed universe, inflation requires that
$\bar S^{1/3}\gta {M_o/T_{\rm col}}$, where
$T_{\rm col}$ must be less than the temperature
at which inflation ensues.
Here we can take $T_c$
to  mean the temperature
at which an inflationary epoch begins.
For example, if
inflation begins at $T_c\sim 10^{14}$ GeV,
then $\bar S\gta 10^{15}$ is needed for the universe
is to survive to $T_c$.
Although the numerical value of $\bar S$ will be
smaller  in an inflationary universe
than in a \mad universe,
the numerical value of $x(T)$ at a given temperature
above $T_{c}$ will be similar.
Comparing $x_{\rm inflation}$ before an inflationary
epoch begins to $x_{\rm mad}$ above temperature $T_c$
shows that
$x_{\rm inflation}=x_{\rm mad}\sim (T_{\rm col}/T)^2$
where $T_{\rm col}$ is chosen less than $T_c$;
that is, $\bar \Omega(T)$ is the same at a given temperature
above $T_c$ in a \mad world as it is before inflation.
The distinction is that the Planck scale in inflation is
only $10^{19}$ GeV while in  \mad it can be many orders
of magnitude larger.
Thus, for $\bar S\sim 1$ the
Planck time
at which $\bar \Omega\sim 1$
would become unstable
is much smaller in a \mad universe
than in inflation.
As a result, larger values of the constant
of motion $\bar S$ are required in the MAD model
to ensure the universe survives until $T_c$.

More generally, if the Planck mass is moving rapidly, the flatness
problem is a bit stickier to discuss although in the end the conclusions
are much the same.  The interested reader is refered to
the appendix.

The flatness problem in an open \mad model has not been discussed here.
We state without proof that the flatness problem persists in the open
model as well. The reason is that the \mad prescription requires
an old universe at a high temperature.  From our experience with the
standard model we learned that if the entropy is of order one,
then the universe would cool below $2.74 ^o$ K in $10^{-11}$ sec.
Similarly, in \mad gravity, if $\bar S\sim 1$, the universe
would rapidly grow cold while the universe was still quite young.

\centerline{\bf Conclusions}

We presented a detailed description of the Brans-Dicke early universe.
For a homogeneous and isotropic cosmology,
the three values of the curvature
$\kappa=+1,-1,0$ separate the Brans-Dicke universe
into expanding and recontracting, expanding forever, and the
critical case between the two extremes,  just as it does with
standard cosmology.

In the Brans-Dicke action there is no coupling of
the Planck mass directly to matter.
As a result, no energy is transfered from the Planck sector
into radiation and the cosmic evolution is adiabatic.
As a direct result of this assumption of adiabaticity, the Brans-Dicke
universe has the usual standard model flatness problem.
An enormous value of the constant entropy $\bar S$ is
required for the universe to survive until today.
However, if a direct coupling of the Planck mass to matter is
considered, then it could be that energy is transfered from the Planck
sector into radiation and entropy is produced.
In the
spirit of inflation,
a large entropy production could explain the present cosmic
flatness.

Any dynamical model which solves the horizon problem automatically
makes
the universe flatter.
For instance in the \mad model, Brans-Dicke gravity
can be used to allow
our present cosmology to be in causal contact during our earliest
history.  In the limit of a slowly rolling
Brans-Dicke field, this is accomplished with a large early value for the
Planck mass and thus weak strength of gravity.
As the strength of gravity increases,
curvature becomes   less and less  important.
Thus the universe does become flatter.
However, because of the large early Planck mass
and thus small Planck time, the universe quickly becomes curvature
dominated
before the strength of gravity increases
unless the universe is very nearly flat at the Planck scale.
As it stands, this generates
the same flatness problem
as in the standard model.
Again, this is a direct consequence of the assumption of
adiabaticity in Brans-Dicke gravity.
The
tenacious flatness problem may encourage us to move away from the
adiabatic assumption and allow for the possibility of entropy
production in a \mad cosmology.$^{10}$

\bigskip
\centerline{\bf Acknowledgements}
\medskip

We offer many thanks to Alan Guth and Alexandre Dolgov for
their thoughts about this project and their
careful reading of sections of the paper.
KF acknowledges support from NSF Grant No. NSF-PHY-92-96020, a Sloan
Foundation fellowship, and a Presidential Young Investigator award.

\centerline{\bf Appendix}

\centerline{\bf MAD Flatness Problem with a Variable Planck Mass}

In this appendix
we will study in some
detail the flatness problem in a closed ($\kappa=+1$) \mad cosmology
with a variable Planck mass.

If $\mu=\dot \p/\p \ne 0$
the flatness problem in a closed
cosmology is  a bit stickier although in
the end the conclusions are much the same.
We work with
the more general collapse temperature
of eqn \helpreview .  We purposely wrote $T_{\rm col}$
to look similar to the collapse temperature in a standard model.
To ensure that the universe survives at least until $T=T_c$,
the temperature at which the causality condition is met,
we can require
that the temperature at which the universe starts to collapse is less
than $T_c$.
Subsequent to
time
$t_c$ the universe will become flatter so we
only have to worry about the very high temperature behavior.

The collapse temperature is clearly more
involved than  if the Planck
mass is constant.
We will study loosely the imposed
requirement that $T_{\rm col}<T_c$ for different
ranges of the constants of integration $S,C,\tilde m_{pl}$
etc.
[We will restrict ourselves to $\om\gta 1$ since we are
using Brans-Dicke gravity for which the observations
have constrained $\om>500$.$^{13}$]
Demanding that $T_{\rm col}<T_c$  gives the requirement
	\eqn\req{{\bar S^{1/3}\left |Q\chi_{\rm col}^{1/2}\right |=\bar
	S^{1/3} \left [ 1+\sqrt{1+Q^2{(4\ep^2-1)\over
	\ep^2}
	}\right ]
	^{1/2}\gta {\p^{1/2}_{\rm col}
	\over T_c\gamma^{1/2}}\ \  ,}}
where $\Phi_{\rm col}$ is defined in eqn \helpagain .

If $Q^2$ is small to moderate, say $Q^2\lta \ $few, then
eqn \req\
reduces to roughly
     \eqn\jimbo{
     \bar S^{1/3}\gta \p_{\rm col}^{1/2}/T_c\ \ . }
Such a large $\bar S$ is consistent with a small $Q^2$
as can be seen from the definition for $Q^2$ in eqn \q .
For a small $Q^2$ eqn \helpagain\ shows
that $\p_{\rm col}\sim \tilde \p$.
Therefore the universe
will begin to collapse when $\p$ nears $\tilde \Phi$.
We can use the causality condition
to constrain $\p_{\rm col}$ and then
make the bound on $\bar S$ more specific.
The weakest requirement on
$\tilde \p$ from the causality condition
came from the slow roll limit of $\p(T_c)\sim \tilde \p$, for which
	\eqn\clash{
	{\tilde \p^{1/2}\over T_c}\gta \beta {M_o\over T_o}
	\  \ .
	}
If the Planck mass had not entered the slow roll limit then
$\tilde \p$ would only have been
driven to even large values than eqn \clash\
demands.  Since $\p_{\rm col}\sim \tilde\p$ here,
we have the bound on $\p_{\rm col}$ of
	\eqn\uha{
     	{\p^{1/2}_{\rm col}\over T_c}\gta
     	\beta {M_o\over T_o}\ \ .
    	}
Finally then
\uha\ in \jimbo\ gives
     \eqn\flato{
      \bar S^{1/3}\gta\beta
	{M_o\over T_o}\ \ .
      }
This is similar to the standard model of cosmology which
needs a very large $\bar S$, corresponding to a nearly flat universe,
to avoid the immediate collapse of the universe.

If
instead $Q^2$ is large then
eqn \req\ becomes roughly
	\eqn\bigq{
	\bar S^{1/3}Q\gta {\p_{\rm col}^{1/2}\over T_c}
	\ \ .
	}
Also, we see from eqn \helpagain , that
      \eqn\dvorjak{
       {\p_{\rm col}\over \tilde \p}\sim Q^{1/2\ep}\ \ .
      }
and $\p_{\rm col}$ is far from $\tilde \p$.
If $Q\rightarrow \infty$
then the curvature dependence is
substantial.
A dominant
curvature drives the Planck scale at which collapse ensues
further and further from the value $\tilde \Phi$.

The causality
condition becomes difficult to satisfy if $Q^2$ is large.
Notice, that at high  temperatures and   values of
$\p$
far from $\tilde \p$, that $\Theta\gg 1$.
Both
a huge $Q^2$ and a huge $\Theta$ suppress the left hand
side of eqn
 \causing\
driving $\p^{1/2}_c=m_{pl}(T_c)$
to higher and higher
scales to reach the demands of this
condition.
Using the causality condition  \causing \
in the constraint eqn \bigq\ gives
    \eqn\whye{\bar S^{1/3}\gta {1\over Q}2\ep
 	\left [{    {\sinh^2\Theta_c +2\ep
	\sinh\Theta_c\cosh\Theta_c+Q^2(1-2\ep)
      \exp({-2\Theta_c})\over
    (\sinh^2\Theta_c+Q^2\exp({-2\Theta_c}))^{1/2}
      }   } \right ]\beta {M_o\over T_o}\ \ .}
 From expression \dvorjak\ we can identify $\exp(\Theta)\sim Q^{1/2}$
and since this is large we can also approximate
$\sinh\Theta\sim \exp(\Theta)/2$.
Putting this information together in \whye\ gives crudely
	\eqn\lastime{
	\bar S\gta {\ep \over Q}\left [{ Q^2(1+2\ep) + Q(1-2\ep)\over
	(Q^2+Q)^{1/2} } \right ]\beta {M_o\over T_o}
	\gta\beta  {M_o\over T_o}
	\ \ }
in the limit of large $Q$.
We find in fact that unless $\bar S$ is large
it is impossible to both satisfy the causality condition
and fix $T_{\rm col}<T_c$.

We conclude in general that although a \mad world gets flatter
below a temperature of $T_c$, the flatness
of the early universe
is not explained.
The huge Planck scale and so very
early Planck time would quickly lead to a curvature
dominated cosmology unless the otherwise arbitrary constant
entropy is quite huge.

\centerline{\bf References}

\item{[1]} C. Brans and C. H. Dicke, {\it Phys. Rev.}
{\bf 24}, 925 (1961).

\item{[2]} D. La and P. J. Steinhardt, {\it Phys. Rev. Lett.}
{\bf 376}, 62 (1989); D. La and P. J. Steinhardt,
{\it Phys. Lett.}, {\bf 220 B}, 375 (1989);
P. J. Steinhardt and F. S. Accetta,
{\it Phys. Rev. Lett.} {\bf 64}, 2740 (1990).

\item{[3]}  A. A. Starobinsky, {\it JETP Lett.}
{\bf 30}, 682 (1979);
A. A. Starobinsky, {\it Phys. Lett.} B {\bf 91}, 99 (1980).

\item{[4]} J. J. Levin and K. Freese, {\it Phys. Rev.} D {\bf 47},
4282 (1993).

\item{[5]} K. Freese and J. J. Levin, unpublished 1992.

\item{[6]} T. Kaluza, {\it Preus. Acad. Wiss.}
{\bf K1}, 966 (1921);
O. Klein, {\it Zeit. Phys.} {\bf 37}, 895 (1926).

\item{[7]} John Barrow, unpublished `Scalar-Tensor
Cosmologies', 1992.

\item{[8]}   K. Freese and J. J. Levin, unpublished 1993.

\item{[9]} A preprint by Y. Hu, M. Turner, and E. Weinberg
also addresses some of these points.

\item{[10]}
The flatness problem is rooted in the notion
that an entropy of $\bar S>10^{90}$ is unnatural.
We have not paused to really ask the question of
what is a natural value for $\bar S$.
Could a huge value of the entropy conceivably
be more natural than a small entropy?

\item{[11]} A. H. Guth, {\it Phys. Rev.} D {\bf 23}, 347 (1981).

\item{[12]} see for instance,
E. W. Kolb and M. S. Turner, {\it The Early Universe}
(Addison-Wesley Publishing Company) (1990).

\item{[13]} See the discussion in C. M. Will,
{\it Theory and Experiment in Gravitational Physics}
(New York:  Cambridge University Press) (1981).

Fig 1:    A schematic picture of the scale factor as a function of
$\Theta$ in a flat Brans-Dicke cosmology.
Time increases along the horizontal axis from right to
left.  The scale factor grows infinitely large as $\Theta$
approaches zero.

Fig 2:    A sketch of the scale factor as a function of $\Theta$ in
an open Brans-Dicke universe.
Time increases from right to left along the horizontal axis.
The scale factor grows infinitely large as $\Theta$ approaches
its minimum value denoted by $\Theta_M$.

Fig 3:    A closed Brans-Dicke cosmology with a growing Planck mass.
This figure shows the development of $\ep^2\mu^2+H_R^2$
versus the development of the absolute value of the curvature term,
$1/R^2$.  For a growing Planck mass ($\mu>0$), the universe collapses
while the square-root $\sqrt{\ep^2\mu^2+H_R^2-1/R^2\ }$is still positive.

Fig 4:    The general behavior of the scale factor as a function
of $\Theta$ in a closed Brans-Dicke universe.  Time increases
from right to left.  The scale factor reaches its maximum extent
at $\Theta_{\rm col}$.  Subsequently the universe begins to shrink.

Fig 5:    A closed Brans-Dicke cosmology with a shrinking Planck mass.
Here is shown a schematic picture of the sum of positive terms,
$\ep^2\mu^2+H_R^2$, versus the magnitude of the curvature term,
$1/R^2$.  As the figure demonstrates, the universe begins to contract
after the sum of positive terms equals the curvature term and the
square root vanishes.  In other words, the universe begins collapse
after the square root goes negative.

\bye

\end